\documentclass[twocolumn,aps,prl,showpacs,floatfix,superscriptaddress]{revtex4}

\usepackage{amsmath}
\usepackage{amsfonts}
\usepackage{amssymb}
\usepackage[dvips]{graphicx}
\usepackage{subfigure}
\usepackage{verbatim}
\usepackage{bm}

\renewcommand{\vec}{\mathbf}
\newcommand{\etal}{\textit{et al.}}

\newcommand{\md}{\mathrm{d}}

\begin{document}

\title{Experimental measurement of acceleration correlations and pressure structure functions in high Reynolds number turbulence}

\author{Haitao Xu}
\affiliation{International Collaboration for Turbulence Research}
\affiliation{Max Planck Institute for Dynamics and Self-Organization,
D-37077 G{\"o}ttingen, Germany}

\author{Nicholas T. Ouellette}
\altaffiliation[Present address: ]{Department of Physics, Haverford College, Haverford, PA 19041, USA}
\affiliation{International Collaboration for Turbulence Research}
\affiliation{Max Planck Institute for Dynamics and Self-Organization,
D-37077 G{\"o}ttingen, Germany}

\author{Dario Vincenzi}
\affiliation{International Collaboration for Turbulence Research}
\affiliation{Max Planck Institute for Dynamics and Self-Organization,
D-37077 G{\"o}ttingen, Germany}

\author{Eberhard Bodenschatz}
\email{eberhard.bodenschatz@ds.mpg.de}
\affiliation{International Collaboration for Turbulence Research}
\affiliation{Max Planck Institute for Dynamics and Self-Organization,
D-37077 G{\"o}ttingen, Germany}
\affiliation{Laboratory of Atomic and Solid State Physics, Cornell University, Ithaca, NY 14853, USA}
\affiliation{Sibley School of Mechanical and Aerospace Engineering, 
Cornell University, Ithaca,  NY 14853, USA}
\affiliation{Inst. for Nonlinear Dynamics, U. G\"ottingen, D-37073 G\"ottingen, Germany}

\begin{abstract} 
We present measurements of fluid particle accelerations in
turbulent water flows between counter-rotating disks using three-dimensional
Lagrangian particle tracking. By simultaneously following multiple particles
with sub-Kolmogorov-time-scale temporal resolution, we measured the spatial
correlation of fluid particle acceleration at Taylor microscale Reynolds
numbers between 200 and 690. We also obtained indirect, non-intrusive
measurements of the Eulerian pressure structure functions by integrating the
acceleration correlations. Our experimental data provide strong support to the
theoretical predictions of the acceleration correlations and the pressure
structure function in isotropic high Reynolds number turbulence by Obukhov and
Yaglom in 1951.  The measured pressure structure functions display K41 scaling
in the inertial range.  
\end{abstract}


\maketitle


Fluid particle acceleration is an important quantity in turbulent flows~\cite{yeung:2002}. For
example, it plays a significant role in the formation of cloud droplets in the
atmosphere~\cite{shaw:2003}. In recent years, advances in
the study of the statistics of acceleration have been made
through the development of Lagrangian experimental 
techniques~\cite{voth:1998,laporta:2001,voth:2002} and the use of 
direct numerical simulations (see, e.g., Refs.~\cite{yeung:2006b,biferale:2004}). 
Using silicon-strip
detectors operating at recording frequencies as high as 70 kHz, La Porta and
co-workers~\cite{laporta:2001,voth:2002} were able to follow passive tracer
particles in a water flow at Taylor microscale Reynolds numbers up to
$R_\lambda \sim 10^3$. Fluid particle accelerations were obtained from the
trajectories of the tracer particles. Their results revealed the highly
intermittent nature of acceleration, and also showed the necessity of
sub-$\tau_\eta$ temporal resolution for obtaining accurate acceleration
measurements, where $\tau_\eta$ is the Kolmogorov time scale, the smallest time
scale in turbulence. In later studies, the same technique was used to
investigate the Lagrangian properties of acceleration following a fluid
particle~\cite{mordant:2004c,crawford:2005}. Due to the one-dimensional nature
of the silicon-strip detectors, however, only one fluid particle could be
followed at a time. Consequently, the spatial properties of acceleration were
not explored in these previous studies.  In other particle tracking
experiments, digital cameras were used to record the motion of tracer particles
and multi-particle statistics were obtained~\cite{ott:2000,luethi:2005}. These
experiments, however, were limited to flows with small Reynolds numbers because
of the slower recording frequency of the cameras. Very recently, advances in
CMOS camera technology have provided the opportunity of measuring the
acceleration of multiple tracer particles simultaneously in high Reynolds
number turbulent flows~\cite{ouellette:2006a}. In this Letter, we present the
first direct experimental measurement of the spatial correlations of
acceleration in turbulent flows with $200 \leq R_\lambda \leq 690$.

Another important quantity in high Reynolds number turbulence that is not
clearly understood is pressure. It has been shown that the clustering of
inertial particles in turbulence is related to the scaling properties of the
pressure field~\cite{bec:2007}. It is, however, extremely difficult to measure
pressure in turbulent flows non-intrusively.  Ould-Rouis
\etal~\cite{ould-rouis:1996} reported that, in the inertial range, the pressure
structure functions computed from the fourth order longitudinal velocity
structure functions scale as predicted by Kolmogorov's K41
theory~\cite{K41,obukhov:1951,moninyaglom:v2} when the Reynolds number is
moderately high ($R_\lambda \geq 230$). However, Hill \&
Boratav~\cite{hill:1997} argued that very large Reynolds numbers are needed to
observe K41 scaling and the assumptions made by Ould-Rouis \etal~result in
large uncertainties in the calculated pressure structure function. Pressure
spectra obtained from numerical simulations~\cite{gotoh:2001} suggested that
the K41 pressure spectrum can only be observed at $R_\lambda > 600$. The
spectra obtained from direct pressure measurements in turbulent jets by Tsuji
\& Ishihara~\cite{tsuji:2003} seem to support this conclusion. In this experiment, 
however, the effect of Taylor's frozen flow hypothesis on the pressure spectra 
has not been fully evaluated.
 
In high Reynolds number turbulence, the acceleration is mostly determined by
the pressure gradient, and the viscous term may be ignored~\cite{yaglom:1949}. 
Under this assumption,
there exist analytical relations between the spatial correlations of acceleration
and the Eulerian pressure structure
function~\cite{obukhov:1951,moninyaglom:v2}. By exploiting such relations, we
obtain an indirect but non-intrusive measurement of the pressure structure
functions in high Reynolds number turbulence.  Our experimental results
strongly support the theoretical predictions of Obukhov \& Yaglom based on K41.



In homogeneous,
isotropic turbulence, the Eulerian pressure structure function depends only
on the separation distance, i.e.,
$\Pi(\vec{r}) \equiv \langle [p(\vec{x}+\vec{r}) - p(\vec{x}) ]^2 \rangle = \Pi(r)$.
For high Reynolds number flows, the fluid acceleration is dominated by the local pressure gradient. Hence,
\begin{equation}
R_{ij}(\vec{r}) \equiv \langle a_i (\vec{x}) a_j (\vec{x}+\vec{r}) \rangle 
= \frac{1}{\rho^2} \bigg \langle \frac{\partial p}{\partial x_i} \bigg\vert_\vec{x} \frac{\partial p}{\partial x_j} \bigg\vert_{\vec{x} + \vec{r}} \bigg \rangle .
\end{equation}
Using the relation between the correlation of pressure gradients and the pressure structure function in homogeneous turbulence, the acceleration correlation can be related to the pressure structure function as~\cite{obukhov:1951}
\begin{equation}
R_{ij}(\vec{r}) = \frac{1}{2\rho^2} \frac{\partial^2 \Pi(\vec{r})}{\partial r_i \partial r_j} .
\end{equation}
In homogeneous, isotropic turbulence, this reduces to
\begin{equation}
\label{eq:Ra}
R_{LL}(r) =  \frac{1}{2\rho^2} \frac{\md^2 \Pi(r)}{\md r^2} , \quad
R_{NN}(r) =  \frac{1}{2\rho^2} \frac{\md \Pi(r)}{r \md r} ,
\end{equation}
where  $R_{LL}(r)$ and $R_{NN}(r)$ are the longitudinal and transverse acceleration correlations, respectively. Therefore, once either $\Pi(\vec{r})$ or $R_{ij}(\vec{r})$ is determined, the other can also be obtained.
It should be emphasized that Eqs.~\eqref{eq:Ra} 
hold in homogeneous, isotropic turbulence at high Reynolds numbers. The
only simplification invoked in deriving these equations is the neglect of a
viscous contribution to acceleration.

In their work, 
Obukhov \& Yaglom further assumed, as first proposed by Millionshchikov~\cite{millionshchikov:1941}
and Heisenberg~\cite{heisenberg:1948a}, that the components of the velocity
gradient are drawn from a multi-dimensional Gaussian distribution. Under this hypothesis,
the pressure structure function in homogeneous, isotropic turbulence satisfies 
\begin{equation}
\label{eq:Dp2}
\frac{\md^4 \Pi(r)}{\md r^4} + \frac{4}{r} \frac{\md^3 \Pi(r)}{\md r^3} = -\rho^2 \Phi(r),
\end{equation}
where $\Phi(r)$ can be written in terms of the derivatives of the longitudinal velocity structure function $D_{LL}(r)$:
\begin{equation}
\Phi(r) \equiv D'_{LL} \bigg(4 D'''_{LL} + \frac{20 D''_{LL}}{r}   + \frac{6D'_{LL}}{r^2} \bigg) + 4 \Big( D''_{LL} \Big)^2.
\end{equation}
An equation for $D_{LL}$ can be obtained from the K{\'a}rm{\'a}n-Howarth equation~\cite{karman:1938} as
\begin{equation}
\label{eq:DLL}
6\nu \frac{\md D_{LL}}{\md r} + \vert S \vert \big[ D_{LL}(r) \big]^{3/2} = \frac{4}{5}\varepsilon r ,
\end{equation}
where $S$ is the structure function skewness~\cite{pope:2000}. Obukhov \& Yaglom~\cite{obukhov:1951} 
 assumed that $S$ is constant for all separations $r$, and so it can be related to the Kolmogorov constant $C_2$ for the structure function $D_{LL}(r)$ as $\vert S \vert = (4/5) C_2^{-3/2} $
and the value of $C_2 = 2.13$ is well known from experiments~\cite{sreenivasan:1995}.
Upon solving Eq.~\eqref{eq:DLL} numerically for $D_{LL}(r)$, Eq.~\eqref{eq:Dp2} 
is solved for $\Pi(r)$ using Green's functions. The acceleration correlations $R_{LL}(r)$ and $R_{NN}(r)$ are then obtained from Eqs.~\eqref{eq:Ra}.


\begin{table*}
\begin{ruledtabular}
\begin{tabular}{cccccccccc}
$R_\lambda$ & $u'$ & $\varepsilon$ & $L$ & $\eta$ & $\tau_\eta$ & $N_f$ & meas. vol. & $\Delta x$ & $N_s$ \\
 & (m/s) & (m$^2$/s$^3$) & (mm) & ($\mu$m) & (ms) & (frames/$\tau_\eta$) & ($\eta^3$) & ($\mu$m/pix) &  \\ \hline 
200 & 0.035 & 7.2$\times$10$^{-4}$ & 61 & 194 & 37 & 37 & 100$\times$100$\times$100 & 80 & 2.5$\times$10$^7$ \\  
350 & 0.11 & 2.0$\times$10$^{-2}$ & 67 & 84 & 7.0 & 35 & 300$\times$300$\times$300 & 50 & 9.0$\times$10$^7$ \\  
460 & 0.25 & 0.28 & 56 & 43 & 1. 9 & 69 & 240$\times$240$\times$240 & 40 & 3.3$\times$10$^7$ \\  
690 & 0.42 & 1.2 & 62 & 30 & 0.90 & 24 & 670$\times$670$\times$670 & 80 & 8.5$\times$10$^7$ \\ 
\end{tabular}
\end{ruledtabular}
\caption{Parameters of the experiments. $u'$ is the root-mean-square velocity. $\varepsilon$ is the turbulent energy dissipation rate per unit mass. $L \equiv u'^3/\varepsilon$ is the integral length scale. $\eta$ and $\tau_\eta$ are the Kolmogorov length and time scales, respectively. $N_f$ is the frame rate of the cameras, in frames per $\tau_\eta$. The measurement volume is nearly a cube in the center of the tank, and its lateral size is given in the units of the Kolmogorov length scale $\eta$. $\Delta x$ is the spatial discretization of the recording system. The spatial uncertainty of the position measurements is roughly 0.1$\Delta x$. $N_s$ is the total number of acceleration measurements.}
\label{tab:exp}
\end{table*}

We carried out three-dimensional Lagrangian Particle Tracking experiments in a
von K{\'a}rm{\'a}n water flow between counter rotating disks. Our experimental
technique and particle-tracking algorithm have been described in detail
previously~\cite{ouellette:2006a,bourgoin:2006,xu:2006a}. Here, we report
measurements from four experiments with Taylor microscale Reynolds numbers
ranging from 200 to 690. The relevant parameters of the flow and the
experiments are shown in Table~\ref{tab:exp}. All measurements were done in the
same apparatus described in Ref.~\cite{voth:2002}, except for the $R_\lambda =
460$ experiment, which was carried out in a new apparatus with a similar
geometry but a different disk propeller. As a result, the integral length scale
$L$ of this experiment is markedly different from $L$ in the other three. In
the $R_\lambda = 460$ experiment, we used the Phantom v7.3 cameras from Vision
Research Inc., which are capable of recording at 37,000 frames per second at a
resolution of 256$\times$256 pixels, nearly a 40\% increase in frame rate
compared to the v7.2 cameras used in the other experiments. Therefore, the
$R_\lambda=460$ experiment has the highest temporal resolution among the four
experiments reported here.

\begin{figure}
\includegraphics[width=0.45\textwidth]{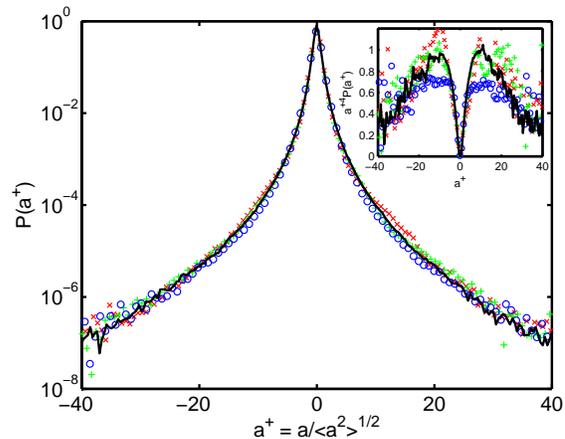}
\caption{(Color online) Measured normalized acceleration $a^+ \equiv a/\langle a^2 \rangle^{1/2}$. The symbols are data from the LPT experiment at $R_\lambda=460$: $\times$ and $+$ are the two measurements of the radial component and $\circ$ are the axial component of acceleration. The solid lines are the previous measurement of the radial component of acceleration using silicon-strip detectors in the same apparatus at $R_\lambda=690$~\cite{mordant:2004}. The inset shows ${a^+}^4P(a^+)$.}
\label{fig:aPDF}
\end{figure}

We have shown before that the probability density functions (PDFs) of
acceleration measured in the $R_\lambda = 690$ experiment agree well with
previous measurements using silicon-strip detectors~\cite{ouellette:2006a}.
Figure~\ref{fig:aPDF} compares the acceleration PDF measured in the $R_\lambda
= 460$ experiment with the PDF measured in Ref.~\cite{mordant:2004} at $R_\lambda = 690$ using
silicon-strip detectors. 
(We note that there is possibly a weak dependence of acceleration PDF on Reynolds number~\cite{voth:2002,chevillard:2003}. 
Previous experiments, however, indicate that the dependence is so small that within experimental uncertainty, the measured acceleration flatness is nearly a constant over the range $285 \leq R_\lambda \leq 970$.)
It can be seen that with a temporal resolution
comparable to the silicon-strip detector measurement, where the sampling
frequency is 65 frames per $\tau_\eta$, the PDFs measured with cameras are in
remarkable agreement with the silicon-strip detector data. Even the fourth
moment agrees well. The spatial resolution in the measurement with cameras, 40
$\mu$m/pixel, is significantly worse than that of the silicon-strip detector
measurement (8 $\mu$m/pixel). The agreement between the two measurements,
however, suggests that using multiple cameras to determine the 3D particle
position results in better accuracy than the one-dimensional silicon-strip
detector measurements.


\begin{figure}
\subfigure[]{
\includegraphics[width=0.45\textwidth]{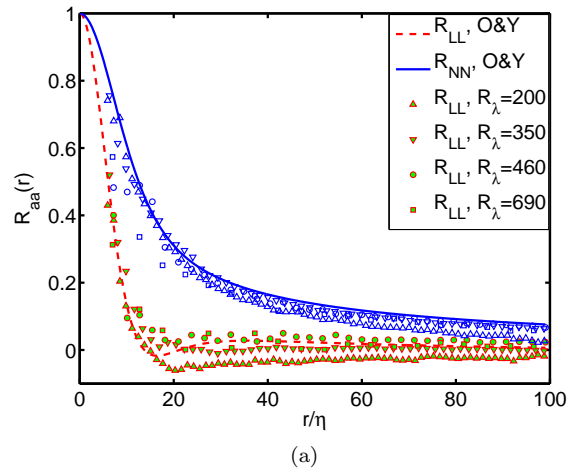}
\label{fig:Raasmallr}
}
\\
\vspace{0pt}
\subfigure[]{
\includegraphics[width=0.45\textwidth]{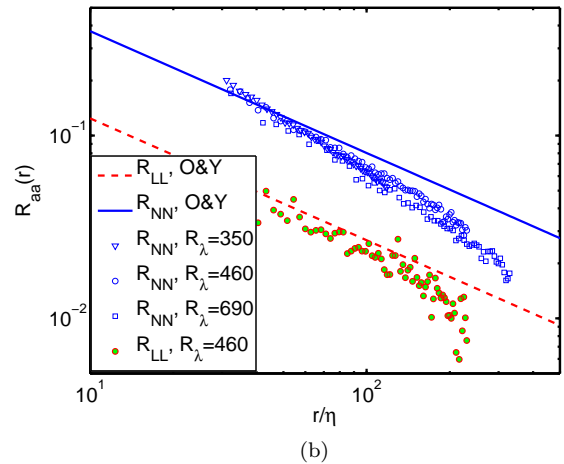}
\label{fig:Raalarger}
}
\caption{(Color online) Comparison of the measured acceleration correlation functions with the Obukhov-Yaglom prediction~\cite{obukhov:1951} for (a) small separations and (b) large separations. The dashed and solid lines are the predicted longitudinal and transverse correlation functions, respectively. The symbols correspond to measurements from LPT data at different Reynolds numbers. Filled symbols are $R_{LL}$ and open symbols are $R_{NN}$.  $\triangle$ -- $R_\lambda = 200$, $\triangledown$ -- $R_\lambda = 350$, $\circ$ -- $R_\lambda=460$, and $\square$ -- $R_\lambda = 690$. }
\label{fig:Raa}
\end{figure}

In Figure~\ref{fig:Raa}, we compare the longitudinal and transverse
acceleration correlation coefficients measured at different Reynolds numbers
with the theoretical predictions by Obukhov \& Yaglom~\cite{obukhov:1951}, as
obtained from Eqs.~\eqref{eq:Ra}. It can be seen from
Fig.~\ref{fig:Raasmallr} that the predictions agree well with the experimental
data. As the Reynolds number increases, the agreement between the predictions
and the measurements increases. This is not surprising given that the viscous
contribution to acceleration is neglected in the theoretical predictions.

At large separations, the predicted acceleration correlations approach simple asymptotic scaling laws~\cite{obukhov:1951}:
\begin{eqnarray}
\label{eq:RaLLlarger}
R_{LL}(r) & \sim & \frac{2C_2^2}{9} \varepsilon^{3/2} \nu^{-1/2} (r/\eta)^{-2/3} , \\
R_{NN}(r) & \sim & \frac{2C_2^2}{3} \varepsilon^{3/2} \nu^{-1/2} (r/\eta)^{-2/3} .
\label{eq:RaNNlarger}
\end{eqnarray}
Figure~\ref{fig:Raalarger} compares Eqs.~\eqref{eq:RaLLlarger} and \eqref{eq:RaNNlarger} with experimental data. There are small but appreciable discrepancies between the prediction and the measurements, which may be caused by the finite measurement volume and/or may reflect the need for still larger separations to see the asymptotic behavior. Another possible reason for the discrepancy is that our flows are not isotropic. We observe anisotropy in acceleration even at the largest Reynolds number investigated, although the anisotropy decreases with increasing Reynolds number~\cite{voth:2002}.

Recently, Hill~\cite{hill:2002} proposed a refined theory for the acceleration
correlations at small separations where the contribution from viscous forces is
not neglected. Due to the limited spatial resolution in our experiment,
however, our measurements cannot be used to test that theory.


We obtain the pressure structure function by numerically integrating
the equation for~$R_{NN}(r)$ in~\eqref{eq:Ra} with experimentally measured transverse acceleration
correlations. As already mentioned before, this equation holds 
in homogeneous, isotropic turbulence at high Reynolds numbers where the viscous contribution vanishes.

\begin{figure}
\includegraphics[width=0.45\textwidth]{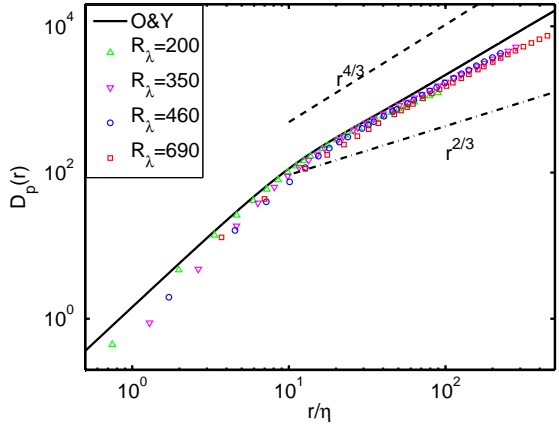}
\caption{(Color online) Comparison of the measured pressure structure functions with the Obukhov-Yaglom prediction~\cite{obukhov:1951}. The solid line is the prediction, and the dashed and dash-dotted lines indicate the $r^{4/3}$ and $r^{2/3}$ power-laws, respectively. The symbols are measurements from LPT data at different Reynolds numbers: $\triangle$ -- $R_\lambda = 200$, $\triangledown$ -- $R_\lambda = 350$, $\circ$ -- $R_\lambda=460$, and $\square$ -- $R_\lambda = 690$.}
\label{fig:Dp2}
\end{figure}

In Fig.~\ref{fig:Dp2}, we compare the prediction by Obukhov and Yaglom [Eqs.~\eqref{eq:Dp2} to~\eqref{eq:DLL}]
with measurements obtained from the acceleration correlations. We plot the normalized pressure structure function $D_p(r) \equiv \Pi(r)/\rho^2 \nu \varepsilon$. The measured data and the predictions are in good agreement over the range of separations accessible in the experiments. K41 inertial range scaling can be obtained from the limiting case of $r \gg \eta$ in the prediction of $\Pi(r)$, yielding
$D_p(r) \sim (r/\eta)^{4/3}$,
which is also plotted in Fig.~\ref{fig:Dp2}. This scaling law is close to the experimental data in the inertial range. The small deviations may be an indication of intermittency. The finite Reynolds numbers and finite measurement volume effects could also contribute to the deviation in the scaling exponent. 

Recently, Bec \etal~\cite{bec:2007} reported a $r^{2/3}$ inertial range scaling for the
pressure structure functions from DNS data up to $R_\lambda = 185$. They also
postulated that the $r^{2/3}$ scaling would persist at low Reynolds numbers and
the K41 scaling might be observed only when $R_\lambda \geq 600$. The $r^{2/3}$
scaling law is shown in Fig.~\ref{fig:Dp2} for reference. As can be seen, all
experimental data are much closer to the K41 scaling rather than $r^{2/3}$
scaling, and there is no appreciable change of slope over the range of $200
\leq R_\lambda \leq 690$. The discrepancies between experimental observation
and numerical simulation remain to be investigated.

Finally, let us note that the extent of the nominal inertial range measured
from the spectrum or from structure functions can be very different, as has
been shown for the case of Lagrangian velocity~\cite{lien:1998}. This subtlety
could account for the difference between our experimental results and previous
investigations, in which the pressure spectra were studied.

In summary, we simultaneously followed the trajectories of multiple passive tracer
particles in turbulent water flows with Reynolds number in the range $200 \leq
R_\lambda \leq 690$. The accuracy of the accelerations measured from the
trajectories is comparable to previous single-particle measurements.  We
obtained spatial acceleration correlations from the multi-particle measurement
and used the measured acceleration correlations to compute pressure structure
functions from a relation that holds at high Reynolds numbers.  We compared
the measurements with theoretical predictions by Obukhov \&
Yaglom~\cite{obukhov:1951} and found that the predictions of both the
acceleration correlations and the pressure structure functions are in good
agreement with the experimental data.  We also observed K41 inertial range
scaling in the measured pressure structure functions over the range of Reynolds
numbers investigated.

\acknowledgments
We thank J\'er\'emie Bec for helpful discussions and Jakob Mann for bringing Ref.~\cite{obukhov:1951} to our attention.
This work was supported by the NSF under Grants PHY-9988755 and PHY-0216406 and by the
Max Planck Society. 

\bibliographystyle{apsrev}

\end{document}